\documentclass[journal=jpccck,manuscript=article]{achemso}
\usepackage{graphicx}
\usepackage[version=3]{mhchem}

\author{Vyacheslav T. Karpukhin}
\email{karp@oivtran.ru}
\author{Mikhail M. Malikov}
\email{mmalikov@oivtran.ru}
\author{Tatyana I. Borodina}
\author{Evgeniy G. Valyano}
\author{Olesya A. Gololobova}
\affiliation
{Laboratory Optical and Physical Measurements, Research Center of Physical and Technical Problems of Power, Joint Institute of High Temperatures (JIHT) Russian Academy of Science, Moscow, 125412, Russia}

\title{Synthesis of Zinc and Zinc Included Nanostructures by High Power Copper Vapor Laser Ablation in Different Surfactants Solutions}

\begin{document}

\begin{abstract}
The data of experimental studies of optical characteristics of colloidal solutions, composition and morphology of its dispersed phase, resulting from laser ablation of zinc in aqueous solutions of anionic surfactants --- sodium dodecyl sulfate (SDS), dioctyl sodium sulfosuccinate (AOT) are presented. It is shown that by studying the optical absorption spectra of the colloid, X-ray spectra and AFM-images of extracted from colloid solid phase, it is possible to trace the dynamics of ZnO nanostructures formation from zinc nanoclasters size of several nanometers to ZnO fractal aggregates (FA) size up to hundreds of nanometers. Determinants of this process are the average power and  an ablation exposure, the frequency of the laser pulses, the colloid aging time, the type and concentration of surfactant in solution. In the selection of appropriate regimes, along with zinc oxide obtained other  nanoproducts --- hydrozincit and organo-inorganic layered composite \ce{[(\beta) - Zn(OH)2 + SDS]}.
\end{abstract}

\section{Introduction}

Over the past decade, among chemical, electro physical and others methods of preparation nanoparticles and nanostructures of  metals, oxides, nitrides,the new technique-pulse laser ablation in liquid environment has demonstrated intense interest~\cite{Yang2007,BozonVerduraz2003}. This method is quite simple, versatile and promotes to have the pure product.A great number of experimental works are devoted for the preparation and investigations of ZnO nanostructures by this method~\cite{Zeng2005,LiYang2007,Usui2005,SunYoungKim2008,Young2006,Zeng2007}. Zinc oxide, semiconductor II-VI group, with a wide band gap $(Eg\approx3.37~eV)$, high exciton binding energy $(\approx60~meV)$ at room temperature is a very interesting material for optoelectronics (UV lasers, diodes, different models of sensors et al)~\cite{Ozgur2005,Willander2009,Shaporev}. The analysis of researches performed on this problem shows that the majority of them were carried out using a low-power Nd:YAG lasers $(\approx1~W)$ with low pulse repetition rate $(\approx10~Hz)$. At the small energy deposition in the ablation process a yield nanoproducts is negligible. The intensification of the process can be achieved by increasing the average laser power output and increase the interaction time with the target material and formed during the ablation colloid, i.e., with increasing of laser pulse repetition rate. In a recent paper~\cite{Wagener2010}, authors using a laser with an average output power $\approx25~W$, picoseconds pulse duration (7~ps) and repetition rate up to 200~kHz, demonstrated the importance  exclusion overlapping zones interaction (bubbles) between successive pulses. It is shown that in the case of overlapping zones interactions, efficiency of production of nanoparticles in this experiment decreased by 30-50 percent and more due to defocusing and scattering radiation incident on the target. At an appropriate choice of spatial and temporal distribution of radiation zones on the target, the performance the synthesis of nanoparticles can be increased at  hundred-thousand times in comparison with the data of the above works with low-power Nd:YAG
lasers~\cite{Zeng2005,LiYang2007,Usui2005,SunYoungKim2008,Young2006,Zeng2007}.
In this regard, the study of the nanostructures synthesis by method  the ablation  different materials in liquids with using high-power lasers with high pulse repetition rate opens up the prospect of effective nanostructures synthesis of  new forms and qualities.

\section{Experimental Section}

For the experiments was chosen copper vapor laser generated pulses of radiation at wavelengths of 510.6~nm ($\approx60\%$ power output) and 578.2~nm with an average power 15-20~W. The pulse width--- 20~ns, repetition rate~--- 10~kHz.The radiation was focused on a target located at the bottom of the glass vessel. The focal length of lens~--- 280~mm, a diameter of spot radiation on a surface zinc ($\approx99.5\%$) target less 100 $\mu$m. The vessel was contained deionized water or aqueous solutions of anionic surfactants~--- sodium dodecyl sulfate (SDS~--- \ce{C12H25SO4Na}) and dioctyl sodium sulfosuccinate (AOT~--- \ce{C20H37NaO7S}). In the case of AOT was used 0.15~M solution AOT in Nonan (\ce{C9H20}~--- organic compound classes of alkanes) followed by dilution with water up to molar ratios water / AOT~--- $W\approx$5, 60, 700. The peculiarity of this solution is the formation of a reverse micelle, the volume of the water pool which depends on W. The volume of the water pool is limited to the size of the nanoparticles~\cite{Tovstun2010}. The volume of solutions was about 2~cm$^3$. The vessel was rotated during ablation time to change laser beam position on a target surface. The solution temperature was measured at a process of ablation. Laser ablation lasted for 30~min and 3~h.

The optical absorption spectra of obtained colloids were recorded after different aging time by UV-Vis spectrometer SF-46 LOMO. The composition and structure of the solid phase colloids studied by X-ray diffraction (XRD; Dron-2, Cu K$\alpha$ radiation). Morphology of the precipitate was analyzed by scanning electron microscopy (SEM; Hitachi S405A, 15~kV)   and atomic force microscopy (AFM; Solver P47-PRO, method of semicontact topography). The samples for these studies were obtained after a centrifugation at 4000 round/min all of colloids and sediments  were washed with deionized water several times to remove main part of surfactant and dried at $\approx60^0$ C temperature.

\section{Results and Discussion}

Figures 1, 2, 3 shows the absorption spectra $A (\lambda$) (in abs. units.) of colloidal solutions obtained for different types of surfactants and their concentrations in the initial solution, the different irradiation time (exposure) $\tau_e$ and aging time $\tau_{st}$ at room colloid temperature. Curves 1, 2, 3, 4 (fig.1) reflect the absorption spectrum of colloid~--- producedby laser ablation of zinc in deionized water.  In all cases the average power irradiation of the target was 12~W.

\begin{figure}[t]
\begin{center}
\includegraphics[width=15cm,draft=false]{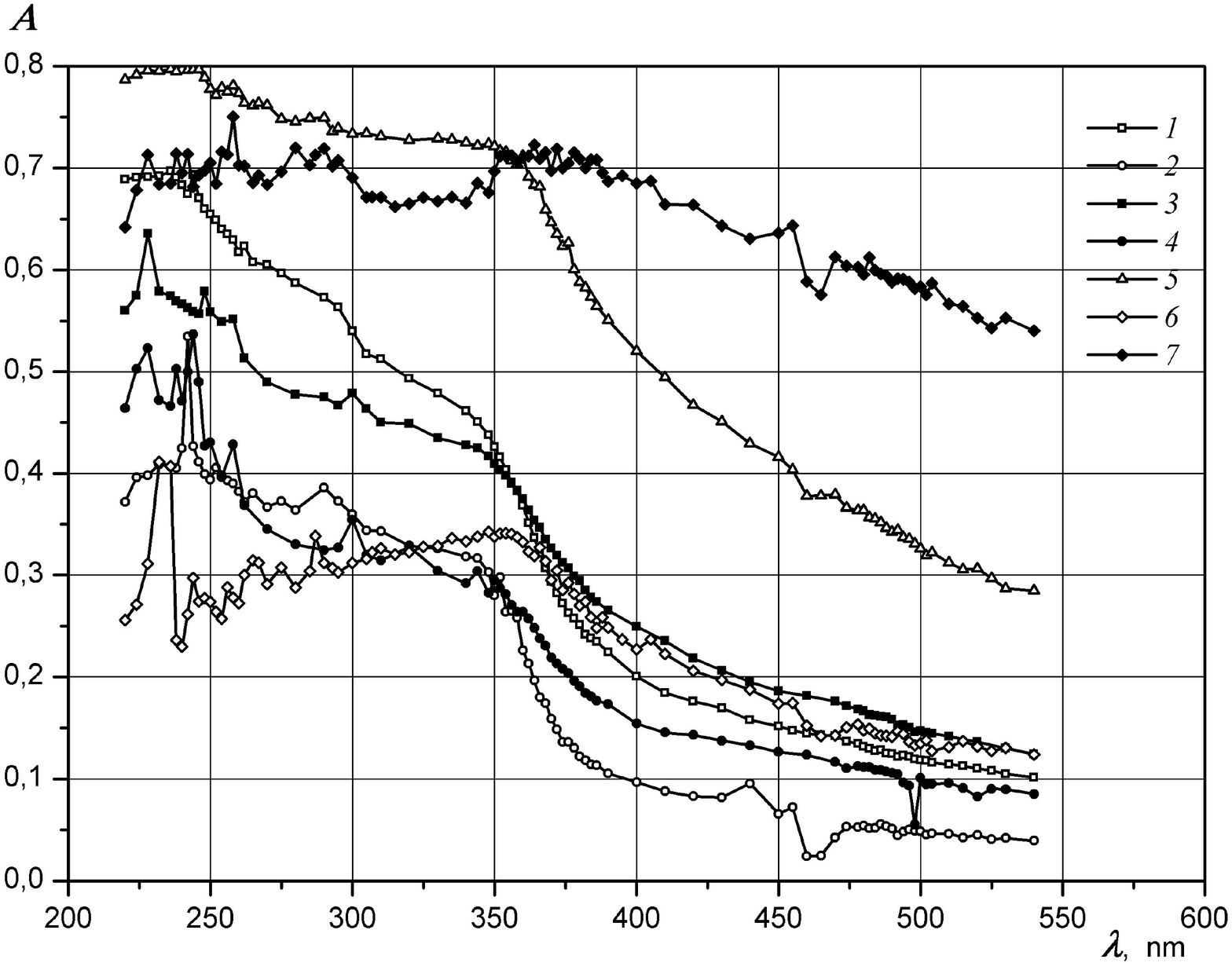}
\end{center}
\caption{The absorption spectra $\rm A (\lambda):  1-Zn + H_2O, \tau_{st}=5~min,\tau_e=35~min;~2-Zn + H_2O,\tau_{st}=3.5~h,\tau_e=35~min;~3-Zn + H_2O, \tau_{st}=15~min, \tau_e=3~h~10~min;~4-Zn + H_2O,\tau_{st}=1~h~40~min,\tau_e=3~h~10~min;~5-Zn + 0.001~Ì SDS, \tau_{st}=7~min,\tau_e=30~min;~ 6-Zn + 0.001~Ì~SDS, \tau_{st}=3.5~h, \tau_e=30~min;~7-Zn + 0.001~Ì SDS, \tau_{st}=30~min, \tau_e=3~h.$}\label{fig1}
\end{figure}

From the analysis, the increase of the surfactant concentration in solution above the critical, defining the beginning of micelle formation in colloid (for SDS is 0.0008~M, and for AOT $<$ 0.01~M) strongly changes the optical characteristics of the colloid. In most cases, in a freshly prepared colloid ($\tau_{st} <$~30~min), the absorption in the UV-range of  spectra ($\lambda <$~350~nm) increases by a half-two times, and in the visible range ($\lambda >$~400 nm) in two or more times ( compare fig.1, fig.2, and fig.3). Maximum absorption in the ultraviolet ( $A>$0.9) is achieved in a solution of 0.15~M AOT in Nonan + \ce{H2O}. A similar result is obtained with increasing exposure time, i.e. with increasing number of nanoparticles of zinc oxide, zinc and other zinc contained products in the colloid. The colloids existence after irradiation for many hours, as a rule, significantly reduces the level of absorption primarily in the longways part of the spectra.

\begin{figure}[t]
\begin{center}
\includegraphics[width=15cm,draft=false]{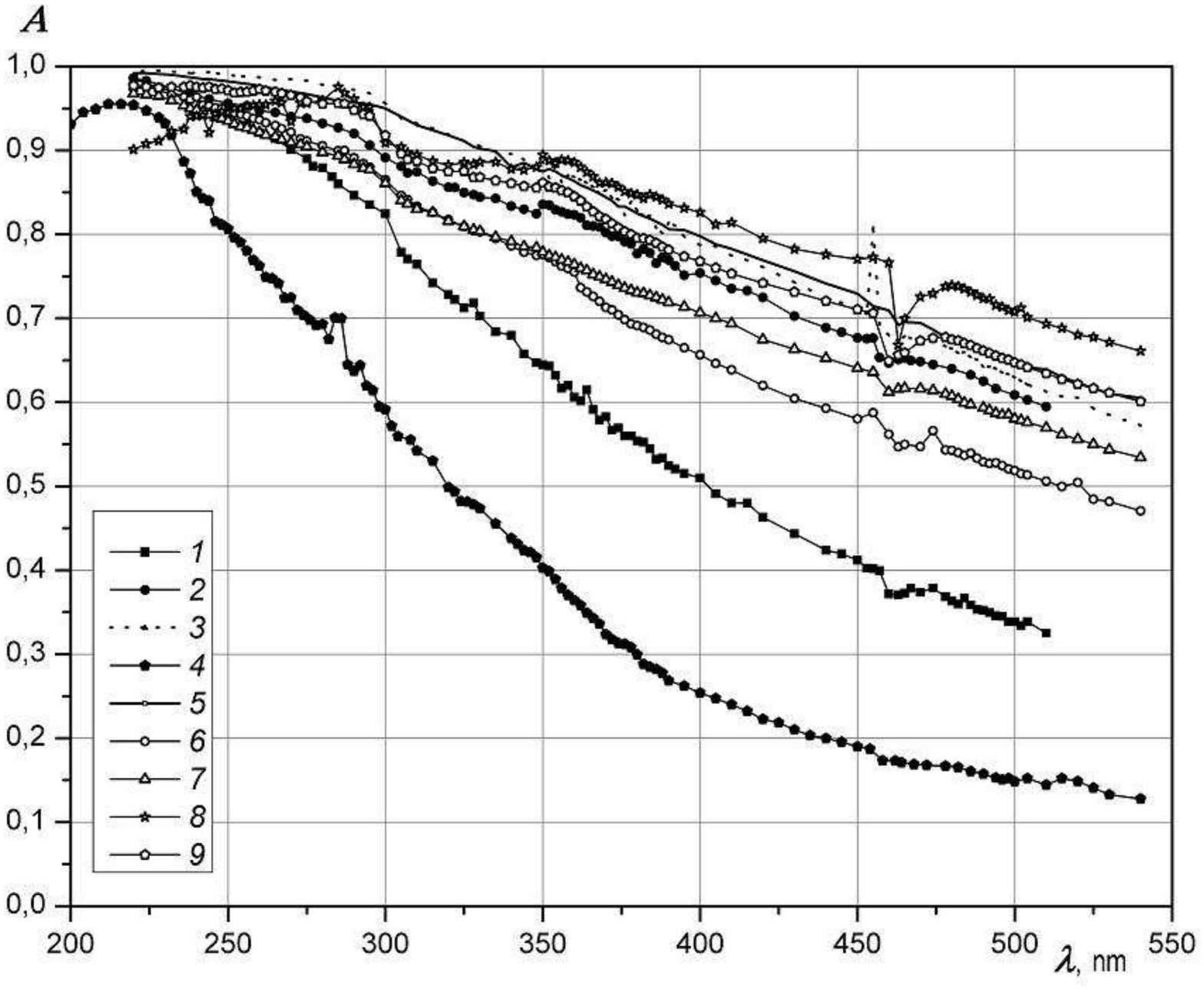}
\end{center}
\caption{The absorption spectra $\rm A (\lambda): 1-Zn + H_2O + 0.01~M~SDS, \tau_{st}=10~min, \tau_e=30~min;~2-Zn + H_2O + 0.01~M~SDS, \tau_{st}=22~h, \tau_e=30~min;~3-Zn + H_2O + 0.01~M~SDS, \tau_{st}=10~min, \tau_e=2~h;~4-Zn + H_2O + 0.01~M~SDS, \tau_{st}=19~h, \tau_e=3~h~16~min;~5-Zn + H_2O + 0.05~M~SDS , \tau_{st}=7~min, \tau_e=30~min;~6-Zn + H_2O + 0.05~M~SDS, \tau_{st}=4.5~h, \tau_e=30~min;~7-Zn + H_2O + 0.05~M~SDS, \tau_{st}=24~h, \tau_e=30~min;~8-Zn + H_2O + 0.05~M~SDS, \tau_{st}=20~min, \tau_e=2~h;~9-Zn + H_2O + 0.05~M~SDS, \tau_{st}=19~h, \tau_e=2~h.$}\label{fig2}
\end{figure}

Peaks of plasmon resonance of zinc nanoparticles ($\lambda$ = 232, 242~nm), visible on the curves in Figure 1 at a case of the zinc ablation in deionized water and for the concentration of SDS in the initial solution 0.001 M (significantly less than the critical micelle concentration - 0.008~M), almost disappear when the molar solution of SDS in water greater than 0.01~M (fig.2). They are not visible, also in absorption spectra of colloids using AOT (0.01~M and 0.037~M) and AOT + Nonan + \ce{H2O} (fig.3).

Well-known exciton absorption band ZnO $(Eg\approx3.37~eV)$ at room temperature, usually determined from the edge of the curve $A~(\lambda$) in the wavelength range 280-380 nm clearly visible only for a cases of zinc ablation in water, SDS solution with concentration of 0.001~M (at the time of exposure $\tau_e = 30$ min) and 0.01~M solution of AOT. The beginning of the band in these cases is about 380~nm. From this it follows that the size of nanoparticles of zinc oxide in the main is more  10-15~nm. In the spectra of other colloids, this band is not observed. In addition, a number of spectra of colloids, the obtained in deionized water, with SDS and AOT, you can see the individual peaks and the characteristic rise of  absorption curves in the wavelength range 280-300~nm, which indicates the presence of zinc oxide particles size 1.5--5~nm~\cite{Pesika2003,Hale2005}.

The peculiarity of the absorption spectra of almost all investigated in this study of colloids is the appearance of a rather narrow ($\triangle\lambda \approx 20$~nm) zone of sharp decrease in the level of absorption (peak of transparency) in the wavelength range 450-480 nm. Peak amplitude increases as with increasing exposure time, and with increasing concentration of SDS and AOT in the initial solution. It is also noteworthy that the peak amplitude decreases significantly with aging time and the peak completely disappears after centrifugation of the colloid fig.2, curve 4, fig.3, curve 2. This means that the appearance of peaks is caused the formation in the solution of fairly larges complexes of particles, which leave it by sedimentation or centrifugation.

\begin{figure}
\begin{center}
\includegraphics[width=10cm,draft=false]{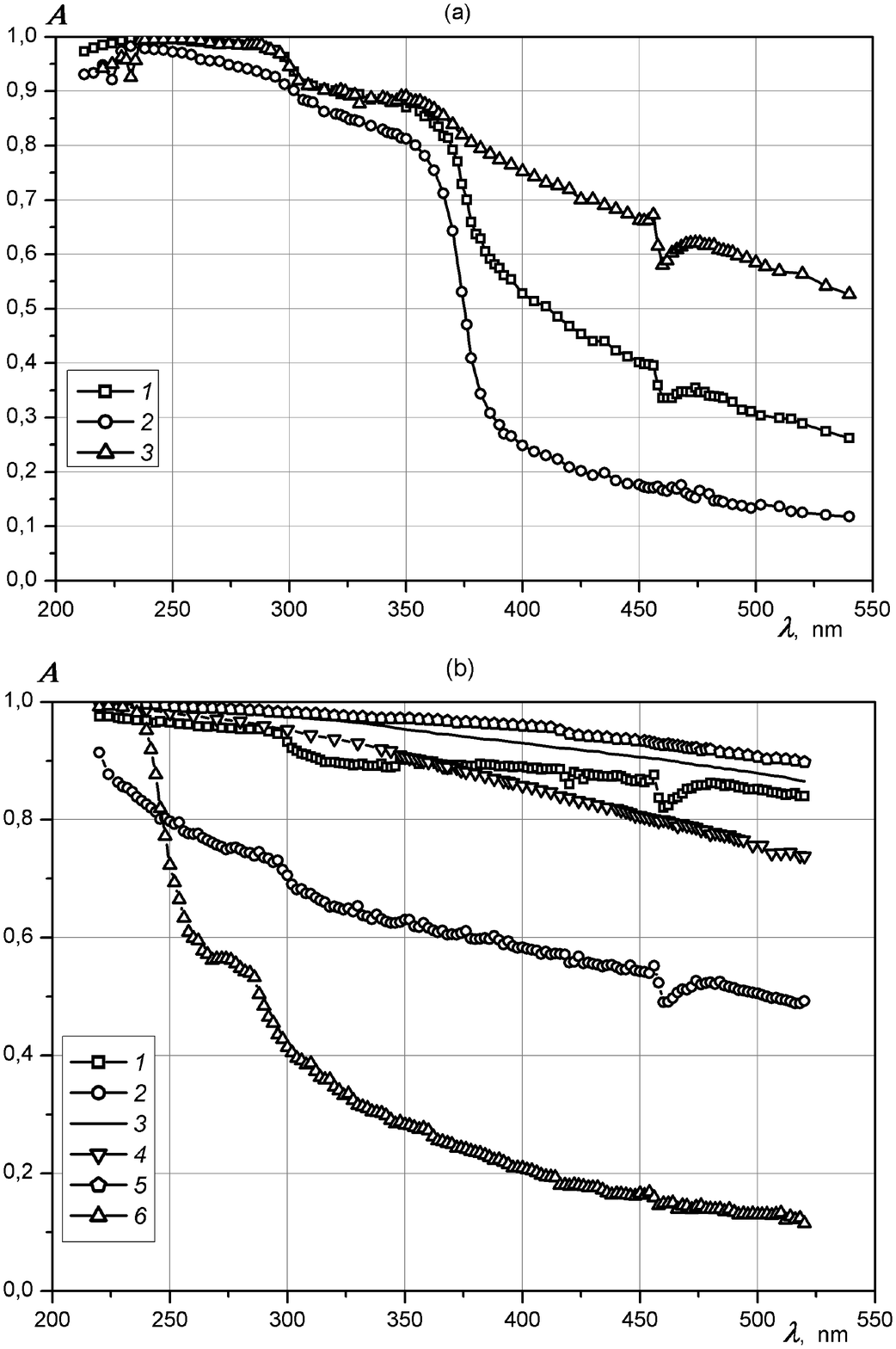}
\end{center}
\caption{The absorption spectra $\rm A (\lambda):(a) 1-Zn + H_2O + 0.05~M~AOT, \tau_{st}=5~min, \tau_e=3~h,~2-Zn + H2O + 0.01~M~AOT, \tau_{st}=22~h, \tau_e=3~h,~3-Zn + H2O + 0.037~M~AOT  \tau_{st}=5~min, \tau_e=3~h; (b)  Zn + H2O + 0.01~M~AOT:~1-W=700, \tau_{st}=30~min, \tau_e=1~h,~2-W=700, \tau_{st}=20~h, \tau_e=1~h,~3-W=60, \tau_{st}=10~min, \tau_e=3~h,~4-W=60, \tau_{st}=4~h, \tau_e=3~h,~5-W=5, \tau_{st}=20~min, \tau_e=3~h,~6-W=5, \tau_{st}=20~h, \tau_e=3~h.$}\label{fig3}
\end{figure}

Nature of the change  the optical spectra is largely confirmed in a number of previously performed studies  nanostructures of zinc oxide, synthesized by a method of laser ablation in liquids~\cite{Zeng2005,LiYang2007,Usui2005,SunYoungKim2008,Young2006,Zeng2007} and a chemical methods~\cite{Shaporev,Hale2005,Kumbhakar}. In different experiments were marked changes of a level of absorption in the UV- and visible part of the absorption spectra~\cite{Young2006,Zang2007}, of  shape and of amplitude of the peaks of plasmon resonances~\cite{Zeng2007}, the shift of  the exciton absorption band of ZnO~\cite{Singh2008,Takahashi2008}.  The presence of narrow peak of transparency is registered in~\cite{Briois2006}.

\begin{figure}
\begin{center}
\includegraphics[width=9cm,draft=false]{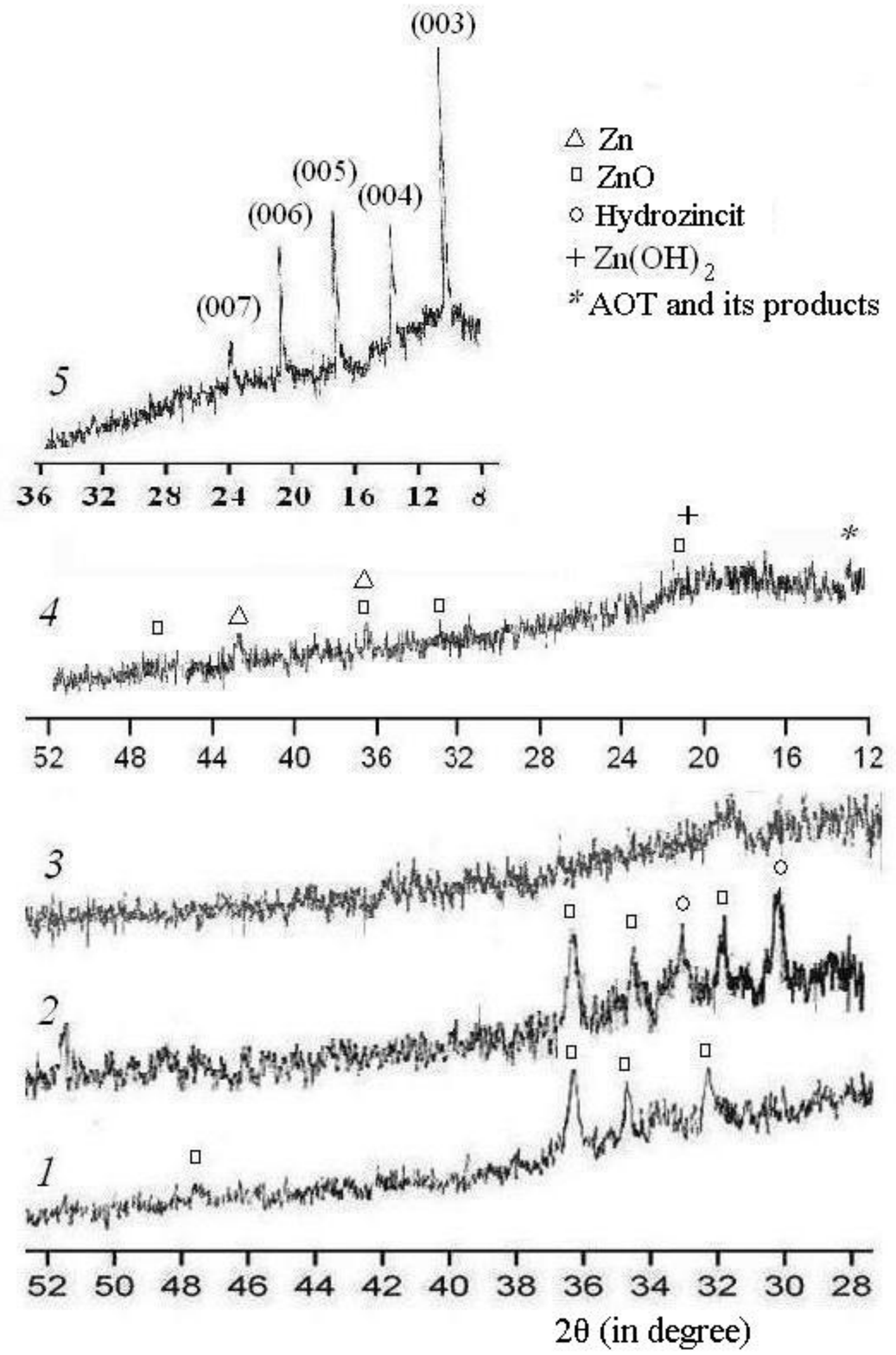}
\end{center}
\caption{X-ray diffraction of sediment extracted from the colloidal solution.~$\rm 1-Zn + H_2O;~2-Zn + H_2O + SDS (M=0.001);~3-M=0.05;~4-Zn + H2O + AOT (M=0.037);~5-Zn + H2O + SDS (M=0.01).$}\label{fig4}
\end{figure}

Interpretation of the absorption spectra of colloids is a difficult task. Indeed, the solid phase colloid consists of products of chemical interaction of zinc with water and surfactant~--- a nanoparticles of zinc and zinc oxide hydrates, organic-inorganic composites produced in the process of ablation and the subsequent aging of the colloid.  All components of the solid phase at different times can change their size and structure due to processes of growth, aggregation and coagulation. Therefore, the overall picture of the optical spectra of a colloids is a superposition of the spectra of plasmon resonances of different nanoparticles, of  peaks and bands of  exciton absorption of zinc oxide,  the absorption bands of other components of  colloids. Nature of the changes in the spectra obtained in the present experiment, their features, the authors believe, can be qualitatively explained in terms of physics of fractals, in particular concerning the optical properties of colloidal systems and composite nanostructures, which are typical examples of fractal systems~\cite{Karpov2003}. The theory of optical properties of fractal aggregates (FA) is different from the well-known Mie theory for metal nanoparticles, since taking into account the electrodynamic interaction of neighbors of the particles that make up the fractal aggregate. Corollaries to this theory which confirmed experimentally are   a broadening and frequency shift of the spectra of plasmon resonances of a fractal, i.e., the increase  of  length of the long-wavelength part spectra in comparison with  a spectra system of individual, non-interacting particles.Frequency shift~---$\triangle\omega-\omega p$, i.e.  a mutual electrodynamics influence of neighboring particles of fractal is so great that it leads to a shift in their resonance on amount comparable to the resonant frequency. The length of the  long-wavelength range of spectra  increases with the size of a fractal. From the theory of optical properties of fractals, in particular, illustrated by the example of silver colloids, it should be possible appearance in the FA spectrum additional peaks and dips in the long-wavelength range of spectra associated with changes in the distribution of particle sizes, including the presence of surfactants in the colloid. There is also a photodynamic effect on colloid particles, the resonances are close to the frequencies of the incident radiation~\cite{Karpov1998}. This effect leads to rather narrow dips~--- "burn-through"~--- in the absorption spectra of the colloid.

In these experiments, the above features of the spectra~--- the rise of the level of absorption especially in the long side of spectra, the emergence of a broad peak in the wavelength range 470-500~nm, the peaks of "transparency" and their disappearance during the sedimentation of large complexes and centrifugation - can be attributed to the provisions of the fractal theory .

Thus, the received optical spectra allow to assume that fractal aggregates of various compositions and shapes are formed in colloids. It proves to be true by X-ray diffraction patterns and an AFM-images of the solid phase extracted from colloids. Fig.4 shows modifications of precipitate X-ray spectra in dependence on a kind and concentration of a surface-active agent in the initial solution. The X-ray spectrum 1 corresponds to the precipitate received after bombarding radiation of zinc in deionized water at $\tau_e=$ 3.5 h and $\tau_{st}=$ 40 h. This spectrum contains accurately expressed peaks belonging to crystalline zinc oxide. This phase has hexagonal crystalline structure with cell parameters $a = 0.3294 \pm 0.0001$~nm, $c = 0.5214 \pm 0.0001$~ nm. The average size of crystallites along the crystallographic axis a is $L_a = 41$~nm, along the axis $c$~--- $L_c = 84$~nm. In this spectrum zinc peaks are not visible.At adding of 0.001~M SDS in the initial solution two crystal phases are exhibited in spectrum 2. One of them~--- ZnO ($a=0.3252 \pm 0.0001$~nm; $c=0.5199 \pm 0.0001$~nm). The crystallites sizes along the crystallographic axes $a$ and $c$ are 40 nm. This is less than in experiment with deionized water. The second crystalline phase is presented in the spectrum by a set of the diffraction lines to the greatest degree coinciding with spectral lines of hydrozincite \ce{Zn5(OH)6(CO3)2}. The average crystallite size of this phase is 40 nm. The hydrozincite amount in the precipitate is rather larger than that of ZnO. Such fact was not registered in known to authors experiments with Nd: YAG-lasers.

Increase of SDS concentration $\rm M = 0.01$ leads to formation of a new phase~--- a lamellar organic-inorganic composition material ZnDS~--- $\rm [(\beta) - Zn(OH)_2 + SDS]$, which is organized by \ce{[(\beta) - Zn(OH)2]} and SDS (the spectrum 5). The similar material has been registered for the first time in~\cite{Liang2004}. The nature of this composition material~--- a chemical bond of hydroxide of zinc \ce{[(\beta) - Zn(OH)2]} with ions SDS. The spectra 3 of the solid precipitate received from the solution with concentration SDS of 0.05~M shows prevalence of an uncrystalline  phase. The crystalline part includes a laminate \ce{[(\beta) - Zn(OH)2 + DS]} and \ce{Zn(OH)2}. Both phases are present in a small amount.

An amorphous substance predominates also in the precipitate after the experiment with the 0.037~M AOT solution. Moreover this precipitate contains crystalline ZnO having the average crystallites size 8~nm, Zn (42~nm) and fine dispersing \ce{Zn(OH)2} (1.5~nm).

The X-ray diffraction studies fulfilled in the present operation and the data of other experiments~\cite{Shaporev,Tovstun2010,Usui2005ultra} show the following dynamics of zinc basis nanoparticles synthesis depending on concentration of anionic surface active agents~--- SDS and AOT. Only Zn and ZnO are formed at ablation in deionized water. At concentrations being essentially smaller than critical value (0.001~M SDS) hydrozincite is received in the conditions of powerful bombarding radiation by means of copper vapor laser. At achievement of SDS concentration about 0.01~M a layered organic-inorganic composition material is formed in experiments with copper vapor laser and at the durable affecting $(\sim 1 h)$ of low-power Nd: YAG-laser radiation~\cite{Usui2006}. The further rise of SDS and AOT concentration leads to formation of finely dispersed yields of zinc ablation. This fact shows that surface-active agents are able to limit nanoparticles growth testifies to developing process of properties. At the same time increase of surfactant concentration in a solution promotes growth and a union of micelles in fractal frames.

\begin{figure}
\begin{center}
\includegraphics[width=7cm,draft=false]{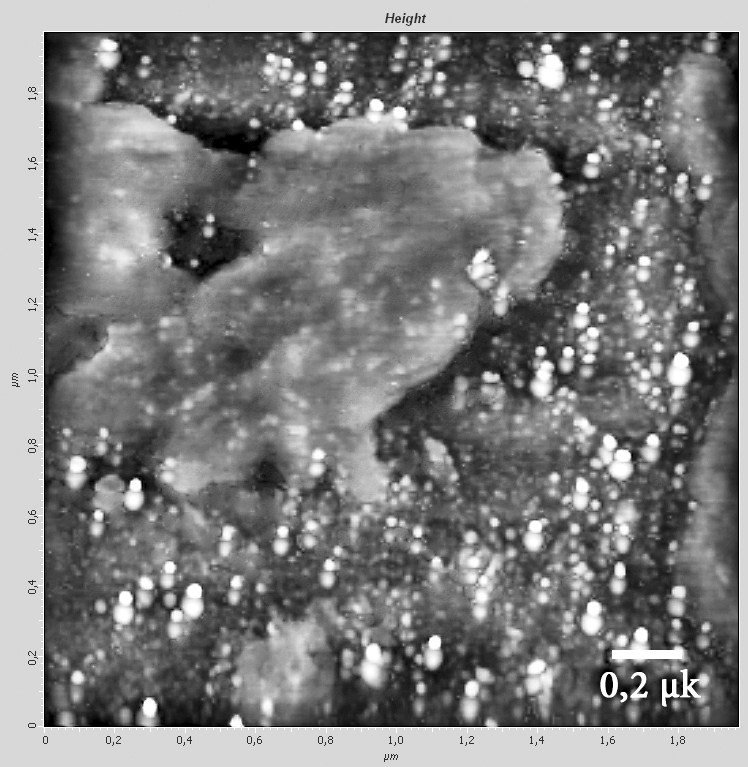}
\end{center}
\caption{AFM-image of the structure of surfaces of the preparation obtained after laser ablation of a target of zinc in the 0.001 M solution of SDS.}\label{fig5}
\end{figure}

Figures 5, 6, 7 present AFM-images of precipitates of the colloids received at irradiation of    solutions about 0.001~M, 0.01~M SDS and 0.037~M AOT accordingly. These pictures confirm the conclusions made on the basis of optical spectrums of colloids and X-ray spectrums of their solid phases. Really, at small SDS concentrations (fig.5) particles in the main of the round shape having sizes of 40-50~nm and less, and also large complexes of the indefinite shape having sizes of 100-400~nm are visible.
\begin{figure}
\begin{center}
\includegraphics[width=15cm,draft=false]{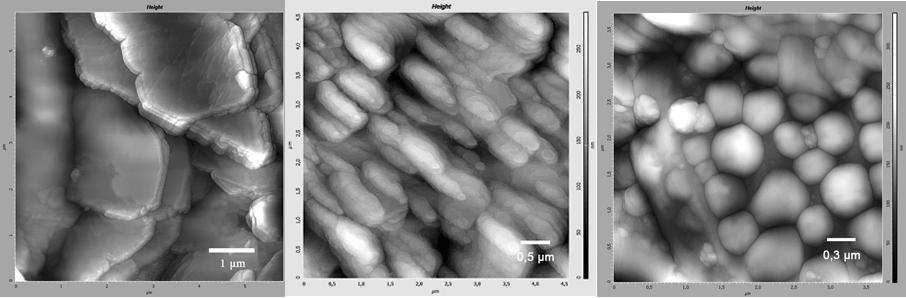}
\end{center}
\caption{AFM-image of the structure of surfaces of the preparation obtained after laser ablation of a target of zinc in the 0.01 M solution of SDS.}\label{fig6}
\end{figure}
On fig.6a, fig.6b it is possible to see colonies (layers) of plates, which are densely laid down against each other and attain in a diameter some microns. On these plates groups of oblate polyhedral particles with round basils are observed (fig.6b). Linear dimensions of these particles lie in the main over the range $\sim$ 200-1000~nm and their thickness crosswise are in 4-6 times less. The similar pattern occurs in samples with AOT (fig.7). Here layered structure with pyramidal microtopography predominates. It is formed by rather large (to 1-3 microns in a diameter) sphenoid flakes. On this flakes particles by linear dimensions $\sim$ 20-150~nm are observed.

\begin{figure}
\begin{center}
\includegraphics[width=7cm,draft=false]{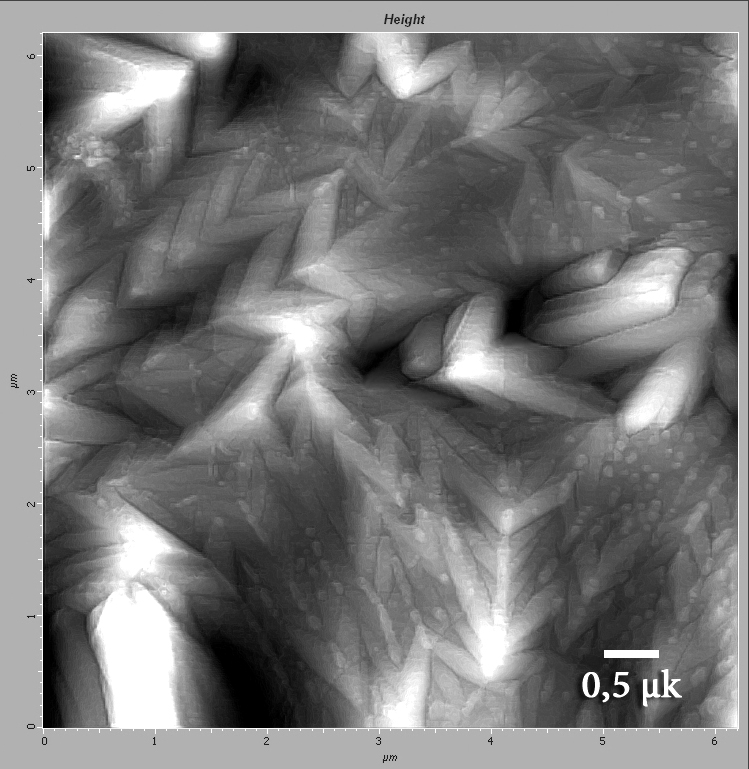}
\end{center}
\caption{AFM-image of the structure of surfaces of the preparation obtained after laser ablation of a target of zinc in the 0.037 M solution of AOT-pyramidal topography formed cuneiform tablets.}\label{fig7}
\end{figure}
\section{Conclusion}

The measured optical spectra, diffraction patterns and AFM-images reflect the dynamics of the formation of nanostructures, resulting from the experiment, a powerful copper vapor laser. The increase in the total time of ablation, the duration of the effects of radiation on the colloid at high repetition rate, rise in temperature of the colloid due to the high average power of irradiation lead to an intensive operating time of nanoparticles of zinc and zinc oxide hydrates, the emergence of clusters and large (up to hundreds of nanometers, and more ) complexes with different structures and forms~---fractal aggregates. Based on fractal theory may explain the qualitative features of the optical spectra of the studied colloids.
Specifics of the use of surfactants in the process of ablation is manifested in the formation of micelles, consisting of surfactant molecules and nanoparticles. On the one hand, the surfactant molecules surrounding the nanoparticle, limit their growth and aggregation, on the other hand, there is a possibility of synthesis of various chemical compounds on the basis of surfactant, water and metal that occurs in these experiments. As follows from literature data obtained organic-inorganic composite \ce{[(\beta) - Zn(OH)2 + DS]} is of interest as a promising equality botchy stochastic environment for UV lasers~\cite{Kalt2010,VanderMolen2010,Markushev2007}.

\end{document}